\documentstyle[sprocl]{article}
\begin{document}
\begin{flushright}
{ITEP-TH-56/96}
\end{flushright}
\vspace*{5mm}
\begin{center}

{\large\bf FUNDAMENTAL CONSTANTS OF NATURE} \footnote{Invited paper to
appear in the Proceedings of the Fifteenth International Conference
on Atomic Physics: Zeeman-effect centenary. Wan der Waals - Zeeman
Laboratory, University of Amsterdam, August 5-9, 1996.}

\vspace{3mm}

{\large L.B.OKUN}\\
{\it ITEP, Moscow, 117218, Russia}

\end{center}

{\small A brief review, from basic atomic constants to "Mendeleev
Table" of leptons, quarks, fundamental bosons, and then to
superunification of all forces and particles.}

\vspace{3mm}

\section{Constants of atomic physics.}

The discovery in 1896 of Pieter Zeeman, which we are celebrating
today, was a great step in unveiling the structure of atoms. At the
same time it was a great step in measuring the fundamental constants
of Physics. As was shown by H.Lorentz, the Zeeman splitting was
determined by the ratio $e/m_e c$, where $e$ and $m$, the charge and
the mass of electron, and $c$, the velocity of light, are three out
of the four fundamental constants of atomic physics. The fourth
constant, $\hbar$, was introduced by Max Planck in 1900. (I am using
the modern notations and terminology.) The fundamental constants
$\hbar, e, m$ are the natural units for atomic physics. They
determine the size and the energy levels of the hydrogen atom (but
not its mass, which, as for any other atom, is determined by the mass
of the nucleus). Three decades later this led to Nonrelativistic
Quantum Mechanics. (An additional important ingredient was spin, the
Pauli principle that explained the Mendeleev Table.)

Already in the original interpretation of Zeeman effect by Lorentz an
important role belonged to the velocity of light, which enters the
expression of the Lorentz force. Were $\hbar, e, m$ the same as they
are, but the velocity of light were infinite, the atoms would not
emit and absorb light, and there would be no Zeeman splitting. In
this sense atomic physics cannot be considered to be
non-relativistic. Note that fine structure constant involves $c$:
$$
\alpha = e^2/4\pi\hbar c
$$

During the XX century the constants $\hbar, c$ took deep roots in
Physics and have fundamentally changed its very basis. The electric
charge $e$ has been joined by the weak and strong charges. As for the
mass of the electron, $m$, it turned out to be one of a whole
constellation of fundamental masses.

\section{QED, leptons and hadrons.}

The Dirac equation combining electron and positron opened a new
chapter of Physics -- the Relativistic Quantum Mechanics, which dealt
with what we now call Feynman tree diagrams. Twenty years later, in
the middle of the century, the Feynman loops became manageable
and the QED (the Quantum Electrodynamics) arose, beautiful, as the
Venus of Botticelli.

But it was clear that this beauty was not alone on the painting. Since
early 1920's protons and since early 1930's neutrons and neutrinos
were known, After the World War II new particles have been
discovered. They belonged to two different groups: leptons and
hadrons. Leptons are: electron, its neutrino and their relatives. The
first leptons, identified after the war were muons. Hadrons are:
proton, neutron and their relatives. The first hadrons discovered
after the war were pions.  Soon they were joined by a crowd of other
strange creatures: strange mesons, hyperons, resonances. Botticelli
was impetuously transforming into a Bosch.

A great relief and order was brought by three ideas:

\begin{enumerate}
\item that all hadrons are particles composed of a few building
blocks (sakatons -- in the 1950's, quarks -- after 1964);

\item that in addition to the electromagnetic interaction, there
are only two other interactions behind all this Boschian chaos: the
strong and the weak one;

\item that the source of strong interaction are three basic, so called
colour charges, whilst the source of weak interaction are two
basic weak charges.
\end{enumerate}

At present the "Mendeleev table" of basic elements consists of 16
particles, not counting antiparticles and colour degrees of freedom
(colour charges).

\section{The "Mendeleev table" of fundamental particles.}

The 16 basic elements are subdivided into two groups: 4 basic bosons
with spin one, and 12 basic fermions with spin 1/2.

The four bosons are carriers of four forces:\\
$\gamma$ -- photon -- of electromagnetic force, with
$\alpha = e^2 /4\pi\hbar c$, \\
$W$ -- of weak force for charged currents, with
$\alpha_W = f^2_W /4\pi\hbar c$, \\
$Z$ -- of weak force for neutral currents, with
$\alpha_Z = f^2_Z /4\pi\hbar c$, \\
$g$ -- gluon -- of strong force, with
$\alpha_s = g^2 /4\pi\hbar c$.

The main difference between photon and $Z$ and $W$ bosons is that
photon is massless, while $m_Z =91$ GeV, $m_W =80$ GeV.

The main difference between photon and gluon is that photon is single
and electrically neutral, while there exist eight gluons carrying
eight different combinations of colour charges, and emitting and
absorbing themselves. The result of this selfinteraction is the
phenomenon of confinement of coloured gluons and quarks inside
snow-white hadrons. The forces between hadrons are not the basic
ones, they are secondary and resemble the Wan der Waals and chemical
forces between atoms.

Twelve fermions are subdivided into three generations -- two quarks
and two leptons in each:

\vspace{3mm}
\begin{center}

\begin{tabular}{|l|c|c|c||c|}
\hline
& 1st & 2nd & 3rd & Q \\ \hline
& $u$ & $c$ & $t$ & 2/3 \\
quarks & & & & \\
& $d$ & $s$ & $b$ & -1/3 \\ \hline
& $\nu_e$ & $\nu_{\mu}$ & $\nu_{\tau}$ & 0 \\
leptons & & & & \\
& $e$ & $\mu$ & $\tau$ & -1 \\
\hline
\end{tabular}

\end{center}
\vspace{3mm}

Each electrically charged fermion has its antiparticle. It may be,
that the same is true for neutrinos, but it is also possible that
neutrinos, like photons, have no antiparticles: each neutrino is its
own antiparticle.
Another unsolved problem, whether neutrinos are massless or have non
vanishing masses.

What are the roles of the three fermionic
generations? The atomic shells are made of electrons, the atomic
nuclei are made of the $u$ and $d$ quarks held together by gluons
inside protons and neutrons: $p=uud$, $n=ddu$. Electronic neutrinos
are needed for weak reactions in the sun and stars. As a result
$$
2e^- +4p \to ~^4He + 2\nu_e +27 \; \mbox{\rm MeV} \;\; .
$$
Without electronic neutrinos there would be no sun and hence we would
not exist. Thus, the first generation of basic fermions is absolutely
necessary for the existence of our world.

The second and third generations seem, at first sight, to be
absolutely useless. But, maybe, they were essential in the first
nanoseconds of the Big Bang by preventing full annihilation of
protons and electrons into neutrinos and photons. Maybe, they had
(and have?) some other functions. They definitely played an important
role in the history of physics. The study of strange particles
(containing $s$-quark) lead to the discovery
of quarks and to the discovery of violation of P, C, CP and T
symmetries in weak interactions, which lead to unification of
electromagnetic and weak interactions into one electroweak
interaction. In accordance with this unified theory (in the Born
approximation, i.e. neglecting electroweak radiative corrections):
$$
\frac{m_W^2}{m_Z^2} = \frac{\alpha_W}{\alpha_Z} =
1-\frac{\alpha}{\alpha_W} \;\; .
$$
One of the key elements of the electroweak theory is the $Z$ boson.
Let us note that the experimental study of $2\cdot 10^7$ $Z$ boson
events at LEP I collider (CERN) has proved that there are only three
light (or massless?) neutrinos. Thus, new particles help to
understand the old ones.

The last free box in the Table of basic fermions has been filled in
only two years ago, when the heaviest quark $t$ was discovered at the
Tevatron collider (FNAL). The mass of this quark is $175 \pm 15$ GeV.

\section{The higgs and the origin of mass.}

It might sound strange, but the value of the top mass is the most
natural one of all leptons and quarks. In order to see this, let us
consider the so called Higgs mechanism, that is used in electroweak
theory to generate masses of fundamental particles. At the basis of
this mechanism lies the (still hypothetical) Higgs field, the
quantum excitations of which are neutral scalar (spinless) bosons --
higgses. The mass of the higgs is unknown at present. In the most
popular scenario higgs is heavier than $Z$ boson but lighter than top
quark. The search for the higgs is the major priority of a new $e^+
e^-$ collider LEP II and of the future Large Hadron Collider (LHC) at
CERN.

Higgs field is coupled to all massive particles, the value of
the coupling constants being proportional to the particles masses.
They are called Yukawa coupling constants.

The unique feature of the Higgs field is that it has a non-vanishing
vacuum expectation value (VEV) $\eta =250$ GeV throughout the world.
Mass of a fermion is a product of its Yukawa coupling times $\eta$.
Masses of $W$ and $Z$ bosons are $g_W \eta/2$ and $g_Z \eta/2$
respectively. The mass of the top quark is the most natural one in
the sense that its Yukawa coupling is of the order of unity.

\section{Running of $\alpha_s$ and confinement.}

For $\eta =0$ all fundamental bosons and fermions would become
massless. This however does not refer to hadrons. Most of them would
remain massive even if the quarks were massless. For instance, the
masses of the proton and neutron would be practically the same, as
they are. This conclusion is deeply connected with the phenomenon of
confinement and with the running of the coupling constant $\alpha_s$.

According to quantum field theory the values of all charges, of all
coupling constants depend on distance (or momentum, or energy). The
constants are changing with these variables because of vacuum
polarization. The famous $\alpha = 1/137.0359895(61)$ is in fact the
value of $\alpha(q^2)$ at a vanishing momentum transfer: $q^2 = 0$. In
the interval from 0 to $m_Z$ $\alpha$ increases from $1/137$ to
$1/129$. The $\alpha_W$ and $\alpha_Z$, in the same interval, change
very little, they "crawl":
$$
\begin{array}{lcl}
\alpha_W(0) = 1/29.01 & , & \alpha_W(m_Z) =1/28.74 \\
\alpha_Z(0) = 1/23.10 & , & \alpha_Z(m_Z) =1/22.91
\end{array}
$$

According to Quantum Chromodynamics (QCD) the behaviour of $\alpha_s$
is totally different; $\alpha_s$ runs in the opposite direction and
runs fast:
$$
\begin{array}{lcl}
\alpha_s(m_Z) \simeq 0.12 & , & \alpha_s(1 GeV) \simeq 1 \;\; ,
\end{array}
$$
and it would "blow up" at smaller momentum transfers, or distances
larger than the radius of confinement, if it were possible to
separate unscreened colour charges by such distances. The
non-perturbative strong self-interaction of gluons, and their
interactions with quarks produces gluon and quark condensates with
characteristic energy scale $\Lambda_{QCD} \approx 300$ MeV. It is
$\Lambda_{QCD}$ that sets the scale of masses of hadrons built from
light quarks ($u, d$) and gluons.

\section{Symmetries and grand unification.}

Up to this point I tried to avoid mentioning symmetries and groups,
using physical, rather than mathematical, language. But in order to
understand the essence of physics one has to appreciate its
mathematical beauty, the beauty of symmetries. First of all, special
relativity is represented by Poincar\'{e} group. Second, QCD is
represented by a local SU(3) colour symmetry with gluons as quanta of
gauge fields of this symmetry. Third, electroweak theory is described
by SU(2)$\times$U(1) gauge symmetry, which is spontaneously broken to
U(1)$_{em}$ by the higgs VEV. Unification of all three types of
interactions is expected to be based on a higher broken gauge symmetry
described by such groups as unitary group SU(5), orthogonal group
SO(10) or exceptional group E$_6$, which contain SU(3) and
SU(2)$\times$U(1) as their subgroups. This idea of grand unification
finds strong support in the fact that the three gauge coupling
constants $\alpha_s$, $\alpha_W$ and $\alpha$ (the latter with a
proper coefficient 8/3), being so different at low energies, tend to
a single meeting point, at $E_{GU}\sim 10^{16}$ GeV, where all of
them have the same value of the order of 1/30.

The fermionic multiplets of higher groups contain both leptons and
quarks. For instance, in the case of SO(10) each generation of
fermions (with account of antiparticles and of three colours of
quarks) forms a 16-plet. Among the 45 vector bosons of SO(10) there
are bosons with such couplings, that their exchange leads to the
proton decay into a positron (or antineutrino) plus accompanying
light hadrons (mesons). Another baryon number violating interaction
produces decays of nuclei, in which two neutrons transform into
mesons; it also transforms neutron into antineutron in vacuum.

The above decays of nuclei have lifetimes longer than $10^{32}$
years, because the corresponding bosons are very heavy: their masses
are of the order of $10^{16}$ GeV. The search for such
decays is one of the highest priorities of the new gigantic
underground detector Super Kamiokande.

\section{SUSY and superstrings.}

A symmetry, which might be broken not so badly, as grand unification
symmetry, is supersymmetry, or SUSY. According to SUSY, there exist
at least one superpartner for each particle we already know. In this
minimal case there exist bosonic analogues of leptons and quarks
(sleptons and squarks with spin 0), and fermionic analogues of bosons
(photino, gluino, zino, wino and higgsino with spin 1/2). The lighter
of these superparticles may be discovered at LEP II and LHC. The
lightest of them might be stable and constitute a substantial part of
the so called dark matter. It is interesting that Feynman loops of
superpartners help to focus more accurately the three running gauge
couplings at the grand unification point.

The energy of grand unification is only four orders lower than the
Planck mass, $m_P$, introduced into physics by Planck, when he
discovered the quantum of action:
$$
m_P = (\frac{\hbar c}{G})^{1/2} = 1.2 \cdot 10^{19} \; \mbox{\rm
GeV} \simeq 2.2 \cdot 10^{-5}\;  \mbox{\rm grams} \;\; ,
$$
where $G$ is the gravitational (Newtonian) constant: $G =
6.6720(41) \cdot 10^{-8} \cdot \mbox{\rm cm}^3 \cdot \mbox{\rm
g}^{-1} \cdot \mbox{\rm sec}^{-2}$. The Planck length, $l_P$, and
Planck time, $t_P$, were introduced in the same paper:
$$
l_P = \frac{\hbar}{m_P c} = 10^{-33} \; \mbox {\rm cm}.
$$
$$
t_P = \frac{\hbar}{m_P c^2} =3 \cdot 10^{-44} \; \mbox {\rm sec}.
$$
At energies of the order of $m_P$, or distances as short as $l_P$, the
energy of gravitational interaction becomes of the order of the total
energy and quantum effects become important. This is the realm of
quantum gravity.

The quantum of excitation of gravitational field is called graviton.
It is massless, neutral and has spin 2. Its source is
the energy-momentum tensor divided by $m_P$. Therefore at low
energies ($E\ll m_P$) its coupling to the matter is extremely weak.
Therefore it has not been observed experimentally, and will not be
observed in the foreseeable future. Even gravitational waves,
classical ensembles of zillions of gravitons, have not been yet
detected by specially built antennas.  But for them prospects are
quite realistic.

A consistent theory of quantum gravity has not been created yet. The
most promising way to it is marked by the sign "superstrings".
Superstrings are tiny one-dimensional objects of the characteristic
Planck length $l_P$, with fermionic and bosonic excitations on them
(therefore the prefix "super"). Most of these excitations are very
heavy, of the order of $m_P$. But there are a few of them which
remain massless. They look like pointlike particles, from distances
much larger than Planckian. Some of the superstring models have
patterns of massless degrees of freedom, which closely resemble some
of the supersymmetric grand unification groups. Thus, superstrings,
are believed not only to provide a selfconsistent theory of quantum
gravity, but to provide it in a broader framework of a unified theory
of all interactions, a theory of everything (TOE). All values of
known (and to be discovered) fundamental gauge and Yukawa coupling
constants are expected to arise as dimensionless elements of
the solution of the TOE equations. It was shown recently that various
superstring models correspond in fact to perturbative expansions in
vicinity of different points of the same theory.

If superstring ideas are correct, then the nature is based on three
fundamental dimensional constants: maximal velocity of particles $c$,
quantum of action and of angular momentum $\hbar$, and Planck length
$l_P$ (or, what is equivalent in units of $\hbar, c$, Planck mass
$m_P$, or Newton constant $G$). The dimensions of other physical
quantities can be expressed in terms of dimensions of $c, \hbar, G$.
In particular, the dimensions of length [L], time [T] and mass [M],
with which elementary physics text-books usually start, are:
$$
[L] = [l_P] \; , \;\; [T] = [t_P] \; , \;\; [M] = [m_P] \;\; .
$$

The $c, G, \hbar$ units has been considered as the most "natural
units of nature" long before the superstrings (Eddington, Gamov,
Ivanenko, Landau, Bronshtein, Zelmanov, Wheeler). From this point of
view the program of Einstein to build a unified theory of gravity and
electromagnetism, without using $\hbar$, was doomed from the
beginning.

\section{Anthropic universe.}

A remarkable feature of our world is how perfectly it is tuned to
favour our existence. The anthropic properties of nature are discussed
in many articles and books. Let me remind a few examples of such fine
tuning in particle and nuclear physics. Start with proton and
neutron. The mass difference $m_n - m_p$ is 1.3 MeV. Were it the case
that this mass difference were 0.5 MeV or smaller, then the neutron
would become stable, whilst the hydrogen atom would be unstable: $e^-
+p \to n + \nu_e$. The most abundant element in the world would be
helium, not hydrogen. The stars would explode at a rather young age.
The genesis of life would become impossible for many reasons.
Analogous dramatic changes are produced by making the electron 0.8
MeV heavier. Note that neutron-proton mass difference is determined
essentially by the mass difference of $d$- and $u$-quarks ($m_d \sim
7$ MeV, $m_u \sim 5$ MeV). Note also that in two other generations
the lower quarks ($s, b$) are not heavier, but substantially lighter
than their upper partners ($c, t$). Compared to the Planck mass, the
tuning of $u$-, $d$-, $e$-masses is of the  order  $10^{-22}$!

Even more striking is the sensitivity of our world to much less
fundamental quantity, such as the binding energy of the deutron,
$\varepsilon = 2.2$ MeV. Decreasing it by only 0.4 MeV would make
impossible the main reaction of hydrogen burning in the sun, $pp \to
d e^+ \nu_e$, so that only the much less effective reaction $ppe^-
\to d\nu_e$ would survive.

Another example is given by energy levels of $^{12}C$ and $^{16}O$.
The famous carbon level at 7.65 MeV lies only 0.3 MeV higher than the
sum of masses of three $\alpha$-particles, and therefore resonantly
enhances the cross-section of the reaction $3\alpha \to ~^{12}C$. The
nucleus $^{8}Be$ being unstable, carbon cannot be produced in two
body $\alpha + ~^{8}Be$ collisions. Without 7.65 MeV resonance the
three-body formation would be not effective enough. As a result
carbon would disappear in the reaction $\alpha + ~^{12}C \to ~^{16}O$
much faster than it would be produced, and the universe would have not
enough carbon to create life.

When looking at the diagram of $^{12}C$ levels (there are about 30 of
them in the interval of 30 MeV) one cannot help admiring that the
level 7.65 MeV does not lie 0.5 MeV lower. The list of such examples
may go on and on. How thin is the margin of safety of everything
which is so dear to our hearts! Most essential features of our world
are determined by absolutely non-essential (from the point of view of
fundamental constants) details of "hadronic chemistry", not speaking
about ordinary chemistry and biochemistry.

The anthropic properties of the universe have led to formulation of a
number of speculative principles.

The weak anthropic principle is based on the notion of an ensemble of
an infinite number of universes with values of dimensionless
fundamental constants, which have been fixed during their
cosmological evolution.
From the very fact of our existence it follows that we
live in one of the best of the worlds.

The cosmological realization of the above statistical ensemble is an
infinite network of universes each of which, at its early
inflationary stage, produces innumerable daughter universes. They may
have different symmetry breaking patterns, even different numbers of
space-time dimensions, and unlimited variety of values of
dimensionless fundamental constants. But here we arrive to the gates
of Metaphysics.

\section{Concluding remarks.}

Looking back at those who made great discoveries at the dawn of our
century and at those who helped them, let us ask ourselves: Was it
possible for any of these pioneers to predict the major steps in
evolution of fundamental physics in the XX century, its impact on the
life of the mankind, and its present landscape? The negative answer
seems to me obvious. It would be even more difficult for us to guess,
what summits the fundamental physics would reach in the next hundred
years, unless the external factors will terminate its development.
Unfortunately, it would be very easy to predict the landscape of
physics and of science in general, if the existing antiscientific
trends would prevail. It will be devastation: intellectual,
scientific, cultural, technological, environmental. The life on our
planet, a unique phenomenon, based on a unique tuning of fundamental
constants of nature, might be ruined. Our duty, as scientists, to be
unanimous and to do our best in defending and promoting fundamental
science.

\section{Bibliography.}

\begin{enumerate}
\item A.Pais. {\bf Inward bound.} Clarendon Press. Oxford. 1986.
\item {\bf Review of Particle Properties.} Phys. Rev. {\bf D50}
(1994) No. 3, part I (an updated edition will appear in 1996).
\item K.Gottfried and V.Weisskopf. {\bf Concepts of Particle
Physics.} v. I,II. Clarendon Press. Oxford. 1984.
\item L.B.Okun. {\bf Particle Physics: The Quest for the Substance
of Substance.} Harwood Academic Publishers. 1985.
\item S.Weinberg. {\bf The First Three  Minutes: A Modern View
of the Origin of the Universe.} Basik Books, Inc. N.Y., 1977.
\item M.B.Green, J.Schwartz, E.Witten. {\bf Superstring Theory.}
Cambridge University Press.
\item A.D.Linde. {\bf Particle Physics and Inflationary Cosmology.}
Gordon and Breach. N.Y. 1990.
\item J.D.Barrow and
F.J.Tipler. {\bf The Anthropic Cosmological Principle.} Clarendon
Press. Oxford. 1986.
\end{enumerate}

\end{document}